# Covariate adjustment for linear models in estimating treatment effects in randomised clinical trials. Some useful theory to guide simulation.

Stephen Senn[1,2,3,*] Franz König[1], Martin Posch[1]
1 Medical University of Vienna, Center for Medical Data Science, Institute of Medical Statistics
2 University of Sheffield, Sheffield Centre for Health and Related Research
3 University of St Andrews, School of Mathematics

*Corresponding author; Stephen Senn, Consultant Statistician, 29 Merchiston Crescent. Edinburgh, EH10 5AJ, UK (stephen@senns.uk)

RUNNING HEAD. Theory to guide simulation of covariate adjustment.

ABSTRACT

Building on key papers that were published in special issues of *Biometrics* in 1957 and 1982 we propose and develop a three-aspect system for evaluating the effect of fitting covariates in the analysis of designed experiments, in particular randomised clinical trials. The three aspects are: first the effect on residual mean square error, second the effect on the variance inflation factor (VIF) and third the effect on second order precision. We concentrate, in particular, on the VIF and highlight not only an existing formula for its expected value based on assuming covariates have a Normal distribution but also develop a formula for its variance. We show how VIFs for categorical variable are related to the chi-square contingency table with rows as treatment and columns as categories. We illustrate the value of these formulae using a randomised clinical trial with five covariates, one of which is binary, and show that both mean and variance formulae predict results well for all $2^5$=32 possible models for each of three forms of simulation, random permutation, sampling from a Normal distribution and bootstrap resampling. Finally, we illustrate how the three-aspect system may be used to address various questions of interest when considering covariate adjustment.

KEYWORDS: analysis of covariance; residual mean square error; second-order precision; simulation; variance inflation factor.

"The fact is that, useful though simulation is, an analytical solution is always preferable …The relative power of an analytic solution, as compared with a simulation approach to a model, is such that even if a full analytical solution is impossible, such a solution to part of the model, with the remainder investigated by simulation, is preferable by far to a simulation solution to the whole model." (Morgan, 1984, P7)

## 1. BASIC MOTIVATION

The choice of our opening quotation is important. It comes from Byron Morgan's classic text, *The Elements of Simulation (Morgan, 1984)*. That a statistician, who has gone to the trouble of writing a monograph on simulation, chooses to stress the importance of analytic solutions, underlines that finding value in simulation as a tool does not mean that theory is not important and *vice versa*. In this paper we try to provide some simple theory regarding analysis of covariance (ANCOVA) that we hope others will find useful.

Much of what we cover is also relevant to adjusting for block effects. This relationship is well covered in an early paper of Finney's in the 1957 issue of *Biometrics* devoted to ANCOVA(Finney, 1957). However, precisely because they are difficult to balance, covariates are harder to deal with than blocks and the distributional theory that is necessary to cover planning for their influence, because random, is not needed when they are blocked. Therefore, we concentrate on the more difficult case of covariates. This difficulty has led many to use simulation as a tool of investigation. Nevertheless, there is some theory that could also be usefully employed but rarely is and it is the purpose of this paper to draw attention to it.

Simulation studies can be valuable for looking at cases where the assumptions under which certain theoretical results have been established break down. Nevertheless, it is frequently difficult to compare and synthesize results across simulation studies because common points of reference are difficult to establish. If theoretical results are also included, these help to anchor simulation studies, making their results more easily understandable but also highlighting precisely what is valuable about simulation studies, namely results for non-standard cases, not easily covered by theory.

A key element of our development is the variance inflation factor (VIF). Although this is a fairly well-known statistic, we believe that many modern investigations of models for adjustment of clinical trials by covariates or blocking factors do not consider it. We propose that paying particular attention to it illuminates many matters including, for example, the relative merits of propensity score (PS) and ANCOVA adjustment. In a previous paper(Senn, König and Posch, 2024), by concentrating on the VIF we were able to discuss the value or otherwise of median stratification by allocation where a single covariate is concerned and a linear model is used. This paper now extends that investigation by looking at multiple covariates.

The outline of this paper is as follows. In section 2 we present in concise form various results that may be used to help in planning clinical trials. Their derivation is given in an Appendix. In section 3 we illustrate the use of these results using an example. In section 4 we give a few examples of questions that may be illuminated with this perspective. Finally, in section 5 we offer a brief discussion. Although we shall cover continuous, binary and categorical covariates, we shall limit our discussion of covariate adjustment to the general linear model and it will be assumed that ordinary least squares will be employed. A further paper in preparation deals with other sorts of model.

Many points we make were made by one or other papers in the special 1957 and 1982 issues of *Biometrics* devoted to ANCOVA, in particular those by Cochran (1957); Cox and McCullagh (1982); Fairfield Smith (1957); Finney (1957). Also important is the paper by Lane and Nelder (1982) regarding the comparison of adjustment in terms of predictions. However, that is particularly relevant for comparing different non-linear models and is beyond the scope of this paper. Unfortunately, this work appears to have often been overlooked in modern examinations of covariate adjustment in randomised clinical trials. One of our purposes is to draw attention to this earlier important work.

As regards the type of covariate we consider, the situation is as described by Fairfield Smith here:

"A postulate underlying the application of analysis of covariance to reduce the effect of extraneous variation on estimated treatment responses is that the concomitant variable, x, is unaffected by treatments, either by direct causation or through correlation with another affected character… The concomitant variable is not necessarily a factor causally affecting the variate… a point, which seems seldom recognized, is that, again assuming the postulate to be true, we need no assumption about the causal relation between x and y. It suffices only that x be correlated with something, often unknown, which causes extraneous variation of y." (Fairfield Smith, 1957, pp284-285).

In other words, we are considering the case of pre-treatment covariates included in the model because they have the potential to increase precision. We aim to cover some simple theory that will help trialists decide whether and to what extent such covariate adjustment will be beneficial

## 2. THE EFFECTS OF COVARIATE ADJUSTMENT ON VARIANCES OF ESTIMATES IN THE LINEAR MODEL

### 2.1. Basic set-up

We restrict our attention to two-armed parallel group designs. In order to develop the theory, we consider the simple case of completely randomised designs. This is not because we regard such designs as common, some form of permuted blocks, for example, is often used(Rosenberger and Lachin, 2016). However, in planning clinical trials, block effects are very rarely accounted for in the accompanying sample size calculations. We discuss this issue briefly in a later section.

Given the set-up described above, then in the case of a linear model with no predictors but two treatments groups with $n_1, n_2$ subjects per group and $N = n_1 + n_2$ patients in total, the variance of the estimated treatment contrast will be

$$\left(\frac{1}{n_1} + \frac{1}{n_2}\right)\sigma_0^2 = \left(\frac{N}{n_1 n_2}\right)\sigma_0^2, \qquad (1)$$

where $\sigma_0^2$ is the within-group variance (assuming homoscedasticity). The subscript 0 indicates that no covariates have been fitted. A lower bound for this is given by the case where $n_1 = n_2 = N/2$ which yields a variance of $4\sigma_0^2/N$.

Compared to (1) there are then three effects on the variance of the contrast of fitting $k \geq 1$ covariates: the effect on the residual mean square error (RMSE), the variance inflation factor (VIF) and second order precision. We shall consider the first and third of these very briefly devoting most attention to the second. However, we shall consider them in order.

### 2.2. Residual mean square error

We assume that we have under consideration, $k$ covariates that are believed to be prognostic of outcome, whatever the treatment given. When fitting these covariates in addition to the treatment factor, the expected RMSE will no longer be $\sigma_0^2$ but $\sigma_k^2 \leq \sigma_0^2$, where $\sigma_k^2$ is the variation in the outcome that is unexplained once the model with $k$ covariates is fitted. As is well-known, for a single covariate, with correlation $\rho$ with the outcome variable, the expected RMSE is $(1-\rho^2)\sigma_0^2$ (Cochran, 1957; Cox and McCullagh, 1982). To the extent that covariates are prognostic, there will be a reduction in the expected RMSE, with equality only applying if the covariates are *not* prognostic, that is to say if $\rho = 0$.

It is sometimes asserted that one will always gain by fitting a prognostic covariate. However, this overlooks the fact that there are two further effects on inference because the parameters are unknown, as discussed below.

### 2.3. Variance inflation

To the extent that covariates are not orthogonal to the treatment indicator, there will be an inflation of the multiplier for the variance of the parameter estimate to be used in conjunction with the residual mean square error(Cox and McCullagh, 1982; Lesaffre and Senn, 2003), so that instead of multiplying the residual variance by $N/(n_1 n_2)$ it will be multiplied by $\lambda N/(n_1 n_2), \lambda \geq 1$, where

$$\lambda = 1/(1-R_Z^2) \qquad (2)$$

and $R_Z^2$ is the coefficient of determination for a regression model in which the covariates are used as linear predictors of the treatment indicator $Z$. Thus $R_Z^2$ is a measure of imbalance (Stewart, 1987) that is conditional on the observed covariate distribution and treatment allocation but can be calculated without knowing the ouctome. When the predictor is a single, categorical variable with $k+1$ categories $k = 1, 2...$, then, as is discussed in the Appendix, an interesting equivalent alternative form is

$$R_Z^2 = Chi^2/N, \qquad (3)$$

where $Chi^2$ is the standard chi-square statistic of association calculated for the $2 \times (k+1)$ contingency table giving numbers of subjects cross-classified by treatment and covariate category.

These forms show a connection to the propensity score, that we shall cover later. $\lambda$ is sometimes referred to as the *variance inflation factor (VIF)* (Stewart, 1987). The value of $\lambda$ will depend on the imbalance across treatment groups of the covariates and can be calculated as

$$\frac{1}{\lambda} = 1 - \frac{n_1 n_2}{N} \mathbf{D}'(\mathbf{X}'\mathbf{X})^{-1} \mathbf{D} \tag{4}$$

where $\mathbf{D}_{k \times 1}$ is the vector of mean differences between treatment groups for the $k$ covariates and $\mathbf{X}_{n \times k}$ is the matrix of the $k$ covariates centred about their respective overall means. See Cox and McCullagh (1982); Senn (2011); Senn, Anisimov and Fedorov (2010) for theory. This form is useful for studying the distribution of the VIF. To simplify further development, we assume that the matrix $\mathbf{X}$ is of rank $k$. However, as indicated by (2), it is really the degree of imbalance of the linear predictor that affects the VIF and this degree of imbalance could be completely determined, even though some individual contributions were not. Consider a two-centre trial with one hospital treating females and the other males but hospitals randomised to give all their patients either treatment or control. Clearly, the effect of *sex* cannot be differentiated from that of *hospital* but the joint effect of both can be eliminated from the treatment estimate.

Note that $\mathbf{D}'(\mathbf{X}'\mathbf{X})^{-1} \mathbf{D}$ is a quadratic form giving a multivariate measure of distance between the two treatment groups and thus also a measure of imbalance. In any given instance, conditioning on the covariates treats them as fixed and the actual calculated value of $\lambda$ is then the penalty that must be paid. This formula does not require the covariates to take on any particular form, for example they can be continuous or binary. However, it can be of interest, when planning a trial and developing the statistical analysis plan, to consider the expected value of $\lambda$. For this purpose, as shown in the Appendix, assuming that the covariates have a multi-Normal distribution and treating the inflation factor, $\lambda$, as a random variable, its expected value is given by

$$E[\lambda] = \frac{N-3}{N-k-3} = 1 + \frac{k}{N-k-3}, \quad N > k+3. \tag{5}$$

This formula can be expressed in terms of the residual degrees of freedom, $\nu = N - 2$, for the model not fitting covariates as,

$$\nu - 1/(\nu - k - 1) \tag{6}$$

and this version can be found on p547 of Cox and McCullagh (1982). These formulae, unlike those given by (2) and (4), require an assumption that the distribution of the covariates can reasonably be modelled as a multivariate Normal distribution. However, we also give reasons for suggesting that even for categorical covariates, of which binary covariates are a special case, it may work well.

This formula clearly indicates that there can be a problem if fitting a large number of covariates in small samples. This is related to the fact that the F distribution does not have a mean if the denominator degrees of freedom are fewer than three (Forbes et al., 2011).

The variance of $\lambda$ (see appendix) is given by

$$Var[\lambda] = \frac{2k(N-3)}{(N-k-3)^2(N-k-5)}, \quad N > k+5. \tag{7}$$

For given fixed $k$, the latter formula, clearly converges to zero as $N$ increases and may be used to plan simulations by considering what the likely Monte Carlo standard errors will be, using $SE = \sqrt{Var(\lambda)/m}$, where $m$ is the number of simulations.

### 2.4. Second order precision

The residual degrees of freedom for estimating the error variance will be reduced by 1 for each covariate fitted. (It is to be understood that a factor with $m$ levels is a set of $m-1$ dummy covariates.) This is a matter of *second order precision* (Siegfried, Senn and Hothorn, 2023). One way of studying this is to look at the effect on critical values of the t-statistic in connection with common levels of significance or confidence (Cochran and Cox, 1957). An alternative approach, not linked to any particular level, is to compare the variance of Students t-distribution to that of the standard Normal distribution. The variance of Student's t with $\nu$ degrees of freedom is $\sigma_t^2 = \nu/(\nu-2)$. Thus, for a two-armed trial in $N$ patients fitting $k$ covariates, we have $\nu = N-2-k$ giving a t-distribution variance of

$$\sigma_{t,N,k}^2 = (N-2-k)/(N-4-k), \tag{8}$$

which can be compared to $\sigma_{t,N,0}^2 = (N-2)/(N-4)$, the case where only the treatment factor but no covariates are fitted.

### 2.5. Accounting for the three factors

Accounting for 2.2, expected RMSE, and 2.3, VIF, as regards the expected effect is relatively straightforward, since they can be regarded as multiplicative and they are both effects on first-order precision (the extent to which estimates will vary randomly). Indeed, such a formula for their joint effect is given by Cox and McCullagh (1982, p547). Including 2.4, is more difficult, since it is a matter of second order precision, affecting the extent to which the RMSE itself will be well estimated. It has long been a matter of controversy. See, for example, the correspondence between Fisher and Nelder (pp 280-283) in Fisher's edited statistical correspondence (Bennett, 1990) and also section 2.3.1 of *Experimental Designs* (Cochran and Cox, 1957) which mentions it and section 74 of *The Design of Experiments* (Fisher, 1990). Fisher's solution, based on fiducial inference, implies that where a treatment estimate has been established with variance $\sigma_{est}^2$ but where that variance is itself based on an estimate with $\nu$ degrees of freedom, the effective precision is not $1/\sigma_{est}^2$ but rather

$$\left(\frac{\nu+1}{\nu+3}\right)\frac{1}{\sigma_{est}^2}. \tag{9}$$

Thus, one might possibly adopt the reciprocal factor

$$\frac{\nu+3}{\nu+1} \tag{10}$$

as a means of calculating the further contribution to a three-factor product of expected RMSE, VIF and 2nd order precision for the purpose of judging the expected effect of covariate adjustment. (Depending on how many covariates were fitted the value of (10) would change.)   For a more modern treatment of the problem in the context of designed experiments, see Gilmour and Trinca (2012). However, we think that studying the three factors separately is, in any case, valuable.

Note also, that the first and third factor are intimately tied up with the linear model. There is no directly analogous effect for other generalised linear models in common use. However, these other models are also affected by the VIF, although the VIF for the non-linear case may be only approximately the same as for the linear case.

### 3. AN EXAMPLE: COVARIATE ADJUSTMENT IN A TRIAL IN ASTHMA

An incomplete blocks cross-over trial in asthma (MTA/02) comparing seven treatments in five periods in 21 sequences is described by Senn et al. (1997). For the purpose of illustrating the effect of fitting covariates, only first period data will be used, thus giving the structure of a parallel group trial, which, for the sake of illustration, will be assumed to be completely randomised. Furthermore, only data from two treatments, placebo and a formulation of formoterol delivered by a particular device at a dose of 24 µg, which will be referred to as ISF24, are retained. One patient has some missing values and their data are not used.

When the data are reduced in this way, values are available for 46 patients. The data will be used to investigate the effect of fitting covariates on the inflation factor $\lambda$, described in section 2. The data are available here: http://www.senns.uk/Data/SJS_Datasets.htm and are given in the dataset labelled "ANCOVA Example". We shall consider the VIF to be expected by fitting five covariates: sex, log $FEV_1$ at baseline, age, height and weight. With the exception of sex, which is binary, all covariates are continuous. A scatterplot matrix for the four continuous covariates is given in **Error! Reference source not found.** and the points have been labelled by the sex of the subject.

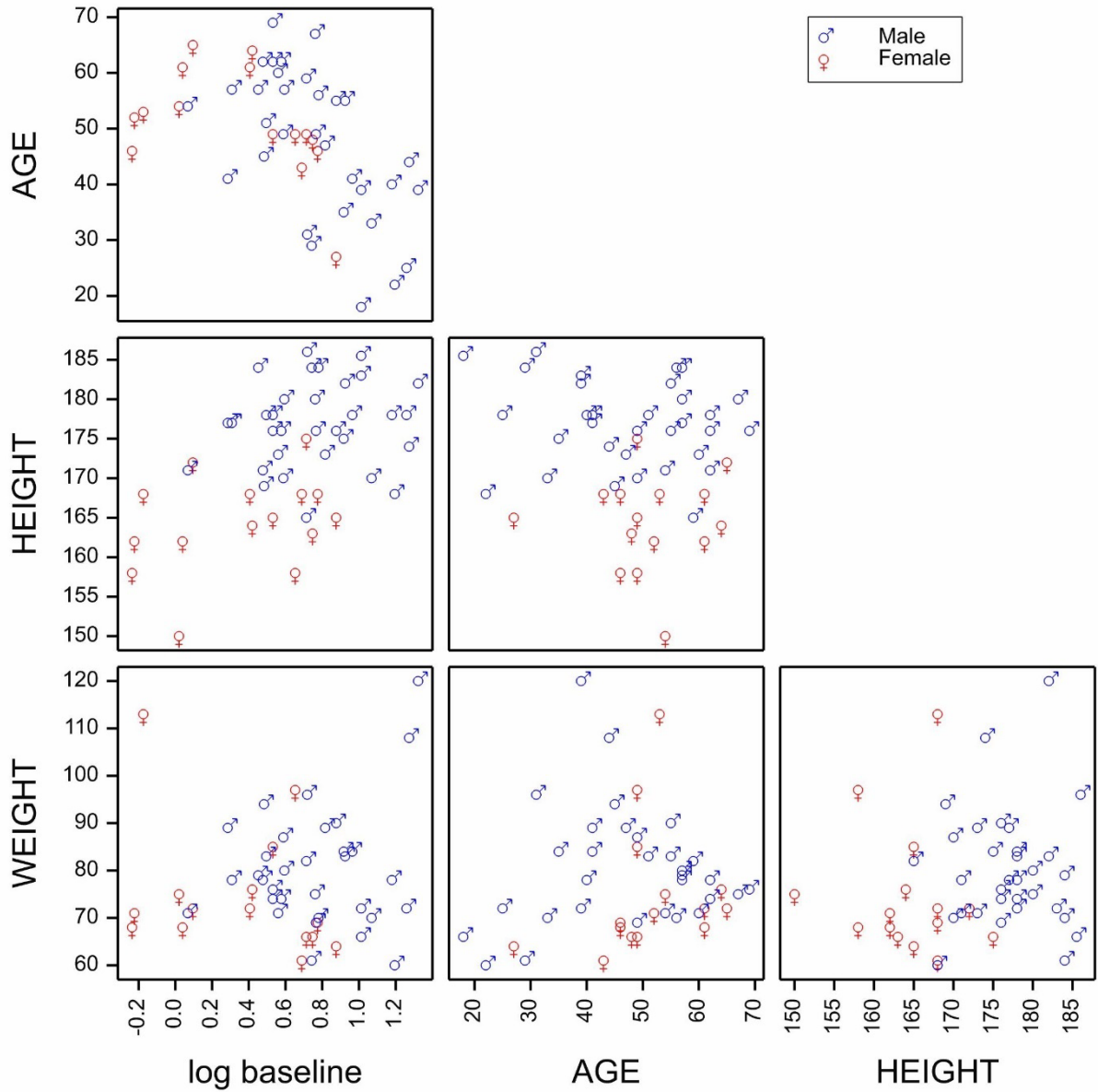

Figure 1 Bivariate scatterplots for the four continuous covariates considered in the example of section 3 with the further covariate sex indicated by symbol.

The details of the simulations carried out in terms of the ADEMP framework of Morris, White and Crowther (2019) are given in Table 1.

| ADEMP term | Description |
|---|---|
| Aims | To illustrate the utility of a formula for expected variance inflation as a function of the number of covariates fitted for a clinical trial, when analysis of covariance is used. |
| Distribution | a) Empirical distribution generated by re-randomisation of the treatment indicators.<br>b) Sampling from a multi-Normal distribution.<br>c) Bootstrap sampling. |
| Estimand | Variance Inflation Factor (VIF) for the given model for the average trial effect. |
| Method | a) Keeping the covariate values fixed for a given patient but choosing the treatment to be allocated using $Bernoulli(½)$.<br>b) Keeping the treatments allocation fixed but randomly generating covariates from a Multinormal distribution with the same mean and variance covariance parameters as the sample statistics for the trial. The binary variable is created by dichotomising a Normally distributed variable at an appropriate cut-point.<br>c) Using sampling with replacement from the 46 sets of covariates in the sample. A set is randomly allocated to each patient, the treatment they were given being retained.<br>For each of the above three cases all possible main-effect covariate models are applied. This yields 32 models in total for each of the three cases as follows:<br>Number of Covariates  0  1  2  3  4  5<br>Number of Models     1  5  10 10 5  1 |
| Performance measures | No particular performance measures for the VIFs are considered but their means and variances are studied using visual inspection. The formula is not expected to work perfectly, since it is based on sampling from a multivariate Normal distribution which will not apply, even for simulation scenario b), since one of the covariates is binary.<br>The empirical mean inflation factors for each scenario for each model are plotted as is the corresponding theoretical value.<br>The empirical variances of the inflation factor are plotted as are the theoretical values. |

*Table 1. A summary of the ADEMP framework for the example. Further explanation is given in the text.*

We now show the results of three sets of simulations. As explained in Table 1, three different sampling frameworks are used. For ease of comparison, all three simulations are represented in the same figures. However, we shall explain each in turn. Within each framework, 32 models are fitted. Half of the models will include sex as a covariate and half will not. Since there is particular interest in the difference between binary and continuous covariates, these two sets are labelled differently. The results of the three simulations investigating the VIF are shown in Figure 1. The horizontal axis gives the number of covariates fitted and the vertical axis gives the VIF. The results are grouped by simulation method as described below. Each point corresponds to one of the $2^5 = 32$ possible models and one of the three possible sampling methods. Thus, there are $3 \times 32 = 96$ points.

The first simulation examines re-randomisation and is labelled *permutation*. The results are represented using the symbol of a circle by the middle set in each triple of sets per number of covariates fitted. For this simulation, the covariate structure is frozen. It is only the treatment allocation that changes. A completely randomised approach is adopted whereby treatment indicators are simulated from a $Bernoulli(\frac{1}{2})$ distribution for each subject. The horizontal dashed lines give the theoretical values that would apply to sampling from a multivariate Normal and fitting 0,1…5 covariates. It is noticeable 1) that despite the fact that sampling from a multivariate Normal does *not* take place, the theory that assumes that it *does* predicts the simulation results extremely well. 2) It appears that it is only the number of covariates that are fitted that matters. The fact, for example, that different pairs of covariates have different degrees of correlation and that one is binary does not appear to matter.

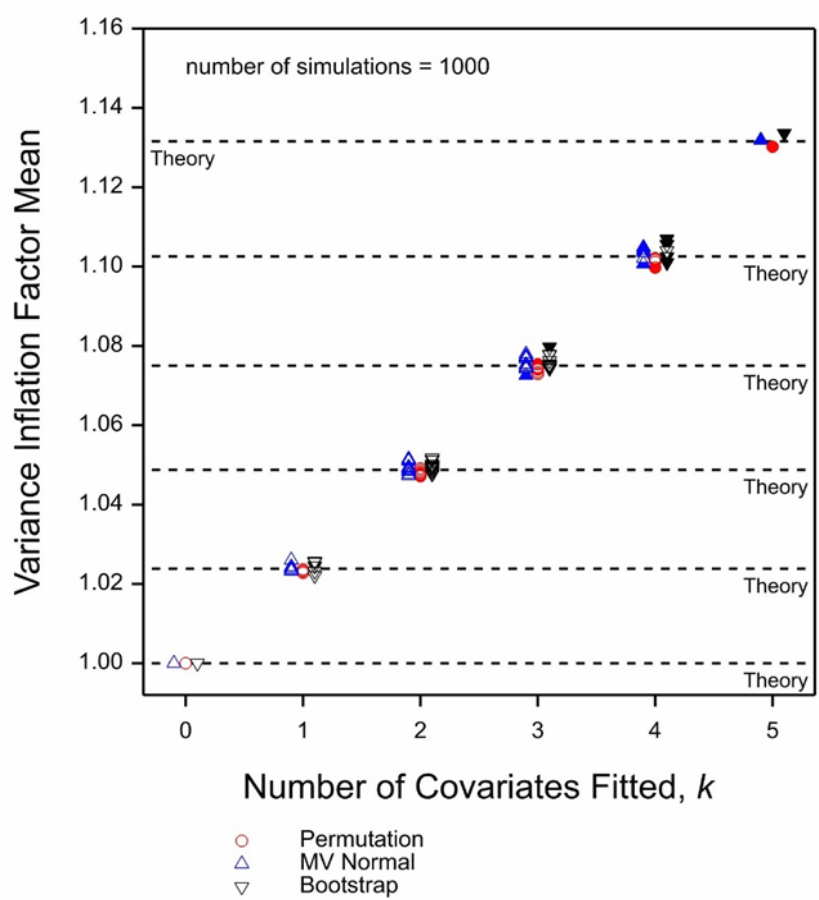

*Figure 1 Simulation to estimate mean VIF for 3 x 32 = 96 analysis of covariance models. See text for explanation. The theoretical values based on 46 patients and 1 to 5 covariates are given by the dashed horizontal lines with 1 covariate at the bottom and 5 at the top. If the symbol is filled, Sex was a covariate in the model. If the symbol was open, Sex was not in the model.*

For each set of three, the left-hand points using an upward triangle give a very similar picture. Here the treatment allocation is fixed but the covariates are simulated from a multivariate Normal with parameters (means, variances and covariance} set equal to the corresponding statistics in the trial. For the covariate *sex*, a Normally distributed value is simulated and then dichotomised to give the same expected distribution as in the trial. (This can be done by coding sex in the dataset as -½ (male) or ½ female, calculating the mean, variance and the relevant covariances of sex, so coded, simulating from the multivariate Normal and then dichotomising at 0.)

In each set of three, the third method of sampling, bootstrap sampling, is shown by the right hand set and indicated with a nabla (downward pointing triangle). The patients are numbered 1 to 46. Random numbers are generated on the interval 0 to 46 and rounded up to the nearest integer. The first patient in the trial is allocated all the covariate values of the patient with the first simulated integer and so on. In this way for a given simulation, some of the covariate patterns will be repeated, some will appear once and some will not appear.

The following points are noticeable. 1) The results are very similar for all three methods of simulation. 2) The results agree closely with the theory, which, however, is developed assuming a multivariate Normal distribution for covariates and thus resembles scenario two most closely. 3) It is only the number of covariates that are fitted that matters.

For any ANCOVA model with a covariate predictor matrix of rank $k$, the matrix of predictors could be replaced by their $k$ principal components. Thus, it is only the rank of the covariates that matters and not their correlation with each other. The only potential surprise, therefore, is that it appears to make no difference, for a given number of covariates, whether the binary covariate *sex* is a member of the set. This point is discussed in the Appendix.

Finally, we can also investigate whether the formula for the variance of the inflation factor holds up well irrespective of the sampling model.

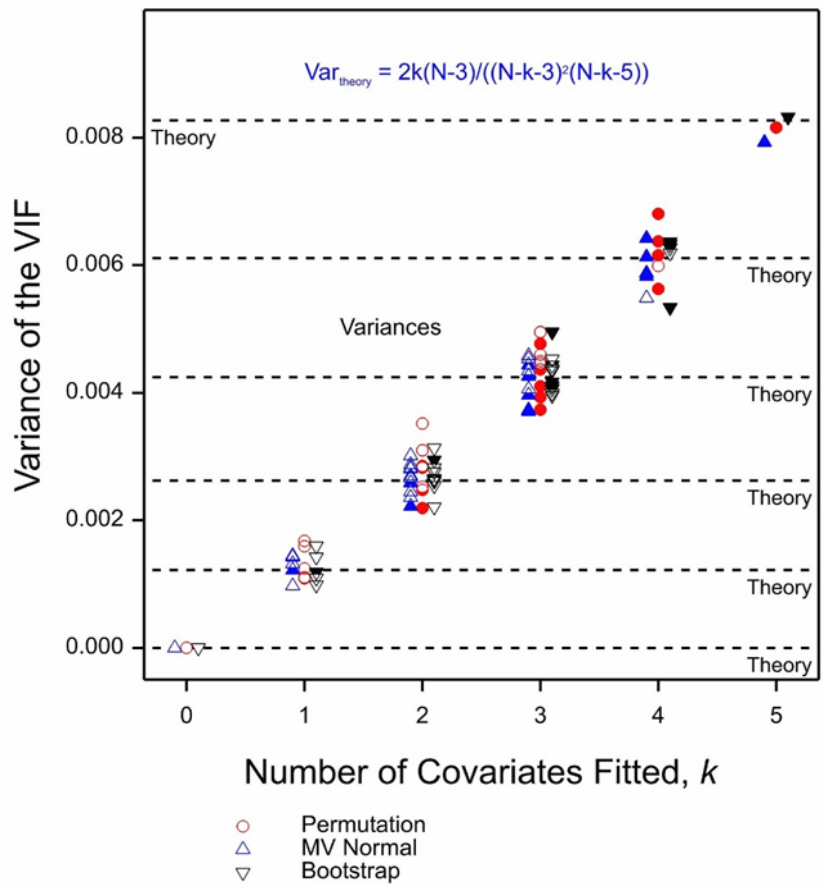

Figure 2 gives the variances of the inflation factors for the 32 possible covariate models for the 3 sampling schemes previously considered. The theory seems to hold up quite well, even though none of the sampling schemes matches theory exactly (given that sex is a binary covariate). If the VIF itself were approximately Normally distributed, theory would suggest that standard errors of its estimated variance based on 1000 simulations would be approximately $\sqrt{2/1000} \approx 0.045$ times the expected value itself. See Stuart and Ord (1993, P364). Obviously, such a figure can only be very approximately applicable in practice.

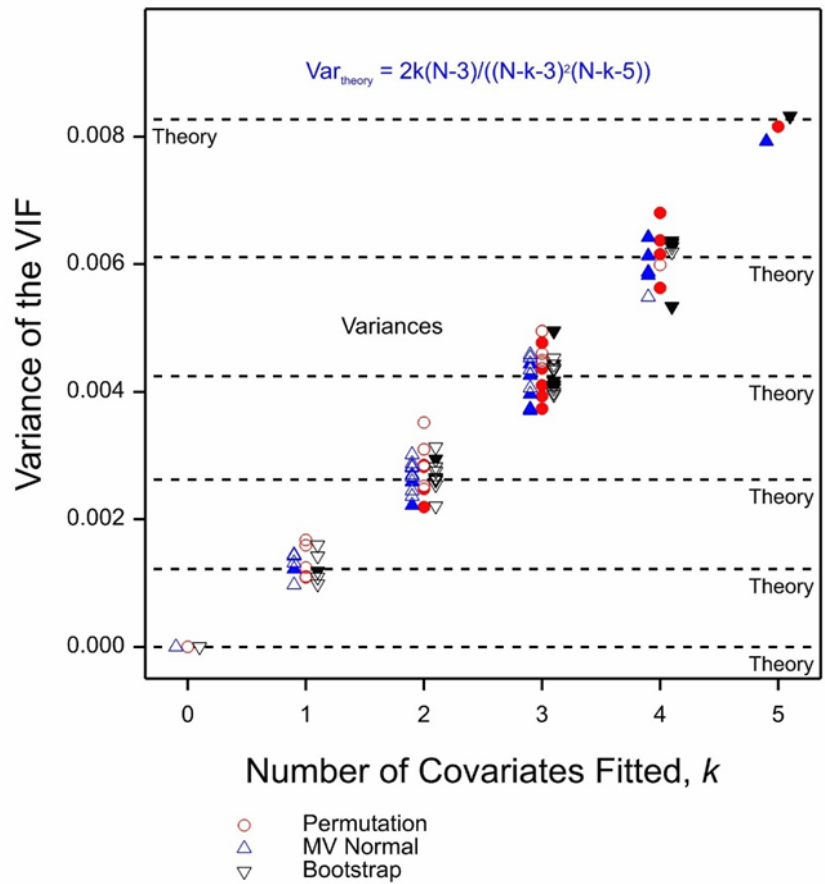

*Figure 2 Simulation to estimate variances of VIFs for 32 analyses of covariance models and the three sampling schemes previously described. If the symbol is filled, Sex was a covariate in the model. If the symbol was open, Sex was not in the model. Given that there were 1000 simulations per point, Normal distribution theory would suggest that standard errors of these variances would be (2/1000) 0.045 times the predicted values.*

Of course, this is only one example and it would be rash to assume that because the theory holds up well here it will generally do so. Nevertheless, we consider that the theory is a useful starting point for any investigation.

## 4. EXAMPLES OF APPLICATIONS

### 4.1. Purpose of this section

In this section we give a few brief examples as to how thinking of a problem in terms of the three factors may help in deciding on a design, choice of covariates to include in a model or even a modelling strategy. The first of these examples is treated in some detail because of central importance. For the rest we restrict ourselves to brief comments.

We make no claim to provide a final analysis of the issues concerned. What we hope to do is show that thinking in terms of our three factors can be a useful start in looking at a problem that may require further examination perhaps involving simulations.

### 4.2. Deciding whether to add a covariate to a model

Consider adding a further covariate to a proposed model in a statistical analysis plan (Gamble et al., 2017) for which $k$ covariates are in the current version then, the ratio of expected VIFs, $r_\lambda$ can be obtained by calculating the ratio of two versions of (5) using $k = k$ for the denominator and substituting $k = k+1$, for the numerator in (5) to give

$$r_\lambda = \frac{N-k-3}{N-k-4} = \frac{v-1}{v-2} = 1 + \frac{1}{v-2}, \quad (11)$$

where $v = N - k - 2$ is the residual degrees of freedom for the model fitting $k$ covariates. The right hand form of 10 is found in Cochran (1957). This factor is always greater than 1. On the other hand, the corresponding ratio, $r_{MSE}$ for expected RMSE is

$$r_{MSE} = 1 - \rho^2_{k+1|k} \quad (12)$$

where $\rho_{k+1|k}$ is the partial correlation of the new covariate with the outcome given others already in the model. This factor is less than one, unless the partial correlation is zero.

The breakeven point in terms of expected true precision of the treatment estimate is given by setting the product of (11) and (12) equal to 1 and is given by

$$\rho_{k+1|k} = \pm \frac{1}{\sqrt{v-1}}. \quad (13)$$

The figure of 0.3 for the absolute value of the partial correlation has previously been suggested as a threshold below which adding the covariate is of little value (Cochran, 1957; Cox and McCullagh, 1982). This yields a value of $r_{MSE} \approx 0.9$. Consideration of (13) shows that this threshold correspond to a value of $v \approx 12$. However, many phase III drug regulatory trials will have hundreds of patients, a study of such trials registered at Clinical-Trials.gov during the period 2008-2019(Brogger-Mikkelsen et al., 2022) found the average number of patients per trial was just over 600, suggesting that it might be possible to fit covariates of much more modest prognostic value with some profit. On the other hand Hee at al found median samples sizes of 74.5, 112 and 122.5 in phase III trials in diseases with prevalences $1 - 9/10^6$, $1 - 9/10^5$ & $1 - 5/10^4$ respectively(Hee et al., 2017; Ristl et al., 2019). Thus, for some trials in rare diseases, the residual degrees of freedom may make it advisable to use models with few covariates. Whatever the practical context, we suggest that considering (13) may be useful.

Of course, this takes no account of second order precision. It is a well known practical point in choice of experimental designs in response surface methodology that some replication be allowed for variance estimation even though such designs are not 'optimal'(Gilmour and Trinca, 2012). A similar issue applies here. Consideration of this point suggests that some further degrees of freedom be allowed for estimating the variance. We speculate that replacing (13) by

$$\rho_{k+1|k} = \pm \frac{1}{\sqrt{v-2}} \quad (14)$$

might be a reasonable rule of thumb.

Alternatively, as already discussed one could adopt and adapt Fisher's approach. We start by using formula (10). If $v$ is the degrees of freedom for the current model, then we can use the formula unchanged. For the model with a further covariate, we must substitute $v-1$ for $v$. The ratio of the two (new to current) will be

$$\frac{(v+2)/v}{(v+3)/(v+1)} = \frac{(v+2)(v+1)}{v(v+3)}. \qquad (15)$$

If the product of this and of (11) and of (12) is set equal to 1 and solved for $\rho_{k+1|k}$ we obtain as break even threshold for the partial correlation

$$\rho_{k+1|k} = \pm\sqrt{\frac{v^2+5v-2}{(v-1)(v+1)(v+2)}}. \qquad (16)$$

The three rules given by (13),(14) and (16) are given in

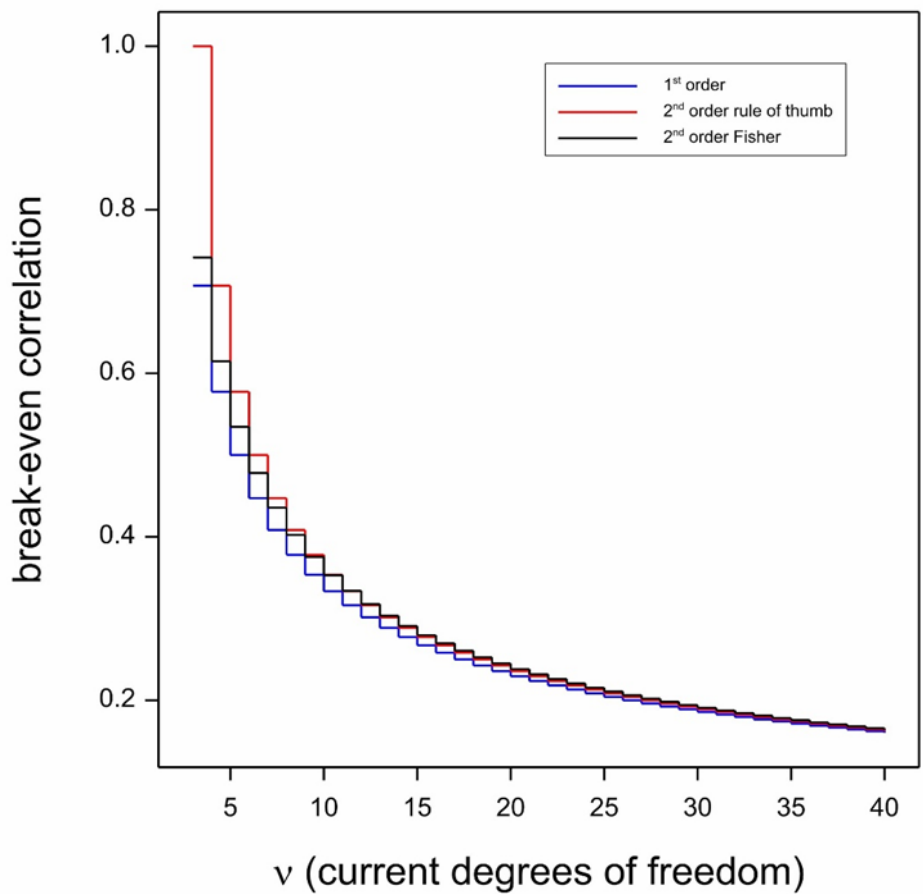

Figure 3. The rule of thumb seems to provide a simple upper bound.

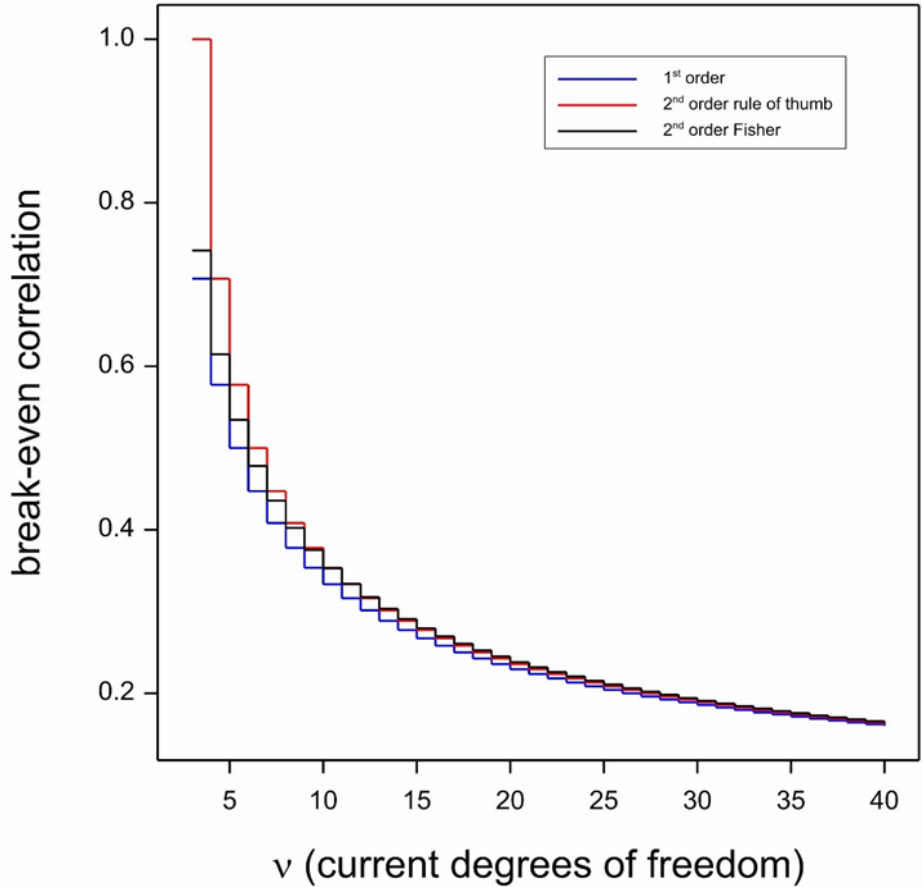

*Figure 3 Three rules for establishing a 'break-even' partial correlation at which it becomes worth fitting an extra covariate.*

The formulae for mean and variance of the VIF clearly show that there is a problem with very small samples in fitting covariates. For typical Phase III clinical trials this is unlikely to be an issue but for other much smaller studies this may not be so. This suggests that reducing the rank of the linear predictor by replacing the individual covariates by a pre-specified score may be beneficial (Siegfried et al., 2023). Where such a score has been established by machine learning from a large historical data-base, it is sometimes referred to as a *super covariate* (Holzhauer and Adewuyi, 2023).

4.3. What is the value of fitting a score based on historical data?

Suppose we have such a historical score based on $k$ covariates (including additional transformed terms of an original covariates) and established using a historical dataset. If we now fit this as our single covariate, one degree of freedom will be used. Alternatively, if we fit instead all the covariates used to construct the score, we shall use $k$ degrees of freedom. From (5) the ratio of expected VIFs is thus given by

$$\frac{N-4}{N-k-3}. \qquad (17)$$

Note that for $k=1$ we have that (17) = 1, which is logical since there can be no advantage to fitting a score based on one covariate. The ratio of expected RMSEs of the two fits will be

$$\frac{1-\rho_c^2}{1-\rho_h^2}, \tag{18}$$

where $\rho_c$ is the correlation coefficient of the linear predictor with the outcome variable having fitted anew using the current data and $\rho_h$ is the corresponding correlation using the historical data. Although (17) is greater than one (provided $k>1$) it is to be expected that (18) will be less than one. We thus have a trade-off.

In conclusion, it is not possible in general to say if fitting historical scores is an advantageous alternative to fitting the covariates anew. It will depend on the sample size. The larger the trial, the less attractive the option of using a historical score. What examining the issue in these terms does suggest, however, is that one should be rather sceptical that using super-covariates based on historical data will make a big difference to the efficiency of phase III clinical trials. It may be inferior to fitting the covariates as a set.

### 4.4. What is the value of minimisation?

Our claim is that there is no value to minimisation(Pocock and Simon, 1975; Taves, 1974) beyond the effect on reducing the VIF. The VIF can never be lower than 1, so that the expected value of the VIF for a completely randomised design permits calculation of the maximum expected reduction that minimisation can achieve.

In practice, expected benefits will be more modest than this for two reasons. The first is that unlike Atkinson's approach(Atkinson, 1982, 2002), VIF reduction is not explicitly targeted. The second is that even where this is targeted the value of 1 cannot, in practice, be achieved. Of course, there is no point in attempting any form of balance of the covariates unless they are assumed predictive and this means that, to estimate the RMSE appropriately, they should be in the model. Failure to do so will lead to an inflation of the expected value of the RMSE estimate compared to a completely randomised design(Finney, 1957).

### 4.5. Categorisation versus smooth functions for modelling awkward covariates

Suppose that for our asthma example, we argue that fitting age as a single linear covariate in the way that we did is naïve and instead propose to create a stratification of age into fifths using the four quintiles and fit it as a categorical covariate. The chi-square test of association between age and treatment allocated is invariant to permutations of the age label. See *Categorical Data Analysis*(Agresti, 1990) section 3.3.4. It is this, see Appendix, which causes one to run the risk of paying a higher VIF penalty. We are thus paying the penalty for some strange forms of imbalance that might not concern us, where, for example, the mean age was very similar in the two groups, although some higher order function was not.

However, suppose that, nevertheless, we insist that we prefer to categorise rather than using simple linear analysis. We can then consider that we have chosen to spend four degrees of freedom in doing this, rather than one, for a possible benefit in terms of RMSE. . We can instead consider that any transformation of age using *up to* four degrees of freedom would pay the same penalty or less in terms VIF and second order precision. The relevant question then becomes, *given the degrees of freedom spent, is the proposed stratification likely to be one that will best explain the variation in outcome*? It seems reasonable that some sort of smooth approach, for example using fractional polynomials (Royston and Altman, 1994; Royston and Sauerbrei, 2008), spending the same degrees of freedom, would do better.

### 4.6. Randomised block designs

Many parallel group trials use some form of blocked allocation. Often this is merely a device to balance numbers either by trial as a whole or, as is often the case, by centre. It is common within the pharmaceutical industry to fit a fixed intercept for each centre and if that is the case, the formulae we have given need some adjustment. One approach is to adapt formula (6) setting $v = N - c - 2$, where $c$ is the number of centres.

In our opinion, statisticians in drug development rarely fit the blocks themselves(Senn, 2000). There are two reasons. One is that they assume that differences between blocks will be small. The other is to avoid spending too many degrees of freedom on eliminating many very small effects. However, if blocks are fitted, the way that (6) can be adapted is obvious.

### 4.7. Choosing between propensity score adjustment and ANCOVA.

Consider the marginalisation formula(Cochran, 1938; Cox and Wermuth, 2003) for the regression of an outcome variable $Y$ on an explanatory variable (such as treatment) $Z$ in the presence of a covariate $X$. This is

$$\beta_{YZ} = \beta_{YZ|X} + \beta_{YX|Z}\beta_{XZ}, \qquad (19)$$

where | is used to indicate conditioning and $\beta_{UV}$ with $U,V = X,Y,Z$ as the case may be, is the regression of $U$ on $V$. There are two sufficient conditions that that $\beta_{YZ} = \beta_{YZ|X}$. The first is that $\beta_{YX|Z} = 0$ and the second is that $\beta_{XZ} = 0$. The first condition is the one that $X$ is not prognostic and the second is that it is 'balanced'.

The propensity score(Rosenbaum and Rubin, 1983) PS stresses factors that are prognostic of assignment . It thus leads one to seek covariates that maximise $R_Z^2$ and therefore, as shown by expression (2) also maximise the VIF Understood like this, it is clear that PS adjustment cannot be expected to produce a lower true variance than ANCOVA provided that a set of true prognostic covariates is available.

| Correlation |           | Sex     | Age     | Baseline | Height  | Weight  |
|-------------|-----------|---------|---------|----------|---------|---------|
| Ordinary    | Outcome   | 0.4197  | -0.5422 | 0.8072   | 0.4328  | 0.0526  |
|             | Treatment | -0.1526 | 0.0167  | 0.0718   | -0.1313 | -0.1254 |
| Partial     | Outcome   | 0.021   | -0.244  | 0.657    | 0.119   | -0.007  |
|             | Treatment | -0.132  | 0.220   | 0.261    | -0.068  | -0.126  |

*Table 2 Ordinary correlations and partial correlations (taking other covariates into account) of the five covariates either with outcome or treatment (the latter being treated as a binary variable). The baseline and outcome FEV1 variables have been log-transformed.*

| Correlation |  | Sex | Age | Baseline | Height | Weight |
|-------------|--|-----|-----|----------|--------|--------|

| Ordinary | Outcome | 0.4197 | -0.5422 | 0.8072 | 0.4328 | 0.0526 |
|---|---|---|---|---|---|---|
| | Treatment | -0.1526 | 0.0167 | 0.0718 | -0.1313 | -0.1254 |
| Partial | Outcome | 0.021 | -0.244 | 0.657 | 0.119 | -0.007 |
| | Treatment | -0.132 | 0.220 | 0.261 | -0.068 | -0.126 |

Table 2 gives ordinary and partial correlations for our example for the five covariates with the outcome variable and the treatment indicator. The correlation with outcome shows (unsurprisingly) that the baseline value is the first candidate to be in the model. However, using 'balance' as a criterion would suggest that sex is far more important. We consider that a philosophy of adjustment that focuses on balance is inappropriate. There are too many matters it cannot address.

Further criticisms of PS adjustment will be found in these papers(Aickin, 2001; Guo and Dawid, 2010; King and Nielsen, 2019; Senn, 2000; Senn, Graf and Caputo, 2007)

### 4.8. Weighting and second order precision

We have mainly considered the effect of VIF and to a lesser extent the RMSE and have said relatively little about second-order precision. For most parallel group trials, the residual degrees of freedom mean that this is relatively unimportant. It is an issue when combining small studies in a meta-analysis because weights based on standard errors are implicitly treated as fixed although they are in fact random. If they all used the same model, weights reflecting sample sizes and VIFs might be a possible solution However, if different numbers of covariates had been fitted. This would penalise trials fitting many covariates because, other things being equal, such trial results ought to be more precise. We thus raise the issue of combining estimates from small trials analysed using ANCOVA as one that may require careful thought.

## 5. DISCUSSION

We hope that we have shown that considerable progress can be made in studying the effect on the variance of treatment estimates of adjusting for covariates and that in doing so, as regards the linear model, it is important to separate three effects: 1. the effect on expected mean square error, 2. the effect on the variance inflation factor and 3. the effect on second order precision via the degrees of freedom for error that remain when fitting the model.

For non-linear cases, factors 1 & 3 are not relevant. However, the variance inflation factor is important and by moving from ordinary least squares to generalised least squares, this can be investigated with the help of formulae for variances for generalised linear and related models. A treatment of this is the subject of a further paper, which will look at generalised linear models.

To return to our main theme, we are not suggesting that simulations are not useful. On the contrary, what we are proposing is that they can become more useful if accompanied by theoretical results, to the degree that this is possible. Where this is so, the theory can provide a partly independent verification of the simulation results and *vice versa*. We say *partly* because, if a conceptual mistake is made in the theory, it may be repeated in the simulation. A simulation may also permit elaboration of a general result from a simple case covered by the theory to a more complex one incorporated in the simulation. However, a theoretical result has the advantage of flexibility. It will usually be easier to make predictions for the whole of the relevant parameter space and even if such predictions are not perfect, they may still be useful, as our opening quotation made clear.

## ACKNOWLEDGEMENTS

We thank Alan Agresti for helpful discussions on categorical variables.

# APPENDIX

## Mean and variance of the inflation factor and connection to the chi-square statistic

*NB. Equation numbering continues that in the main paper in order to facilitate cross-referencing.*

The formula for the expected mean inflation that we present here and is not new and is given, for example, by Cox and McCullagh (1982). However, the derivation is useful for understanding on what it depends and also permits us to go further and derive also the variance formula given by (7).

We exploit the connection between linear two-group discriminant analysis using regression with a binary outcome for group membership and comparison of two groups in terms of multivariate analysis, say using Hotelling's $T^2$ test. The former conditions on the covariates and the latter treats them as random and having a multivariate Normal distribution. A useful discussion of the relationship is found in the book by Flury (1997), in particular sections 6.4 and 6.5 .

From Flury's equation (17) on page 418 we have

$$R_Z^2 = \frac{n_1 n_2 D_{MV}^2}{(N)(N-2) + (n_1 n_2 D_{MV}^2)} \tag{20}$$

where $R_Z^R$ is the coefficient of determination obtained by regressing the treatment indicator on the covariates and $D_{MV}^2$ is the (squared) multivariate distance. However, from P405 we have that

$$\frac{N-k-1}{(N-2)k} \frac{n_1 n_2}{N} D_{MV}^2 = F_{Rao} \tag{21}$$

is equal to what is sometimes referred to as Rao's F-statistic (Rao, 1948) and has an $F(k, N-k-1)$ distribution. Then, using (2) we obtain that the variance inflation factor is related to Rao's F-statistic by

$$\lambda = 1 + \frac{k}{N-k-1} F_{Rao}. \tag{22}$$

However, in general the mean of a random variable that is distributed $F(v, \omega)$ is

$$E\left[F_{v,\omega}\right] = \omega/(\omega - 2), \quad \omega > 2 \tag{23}$$

and its variance is

$$V\left[F_{v,\omega}\right] = \frac{2\omega^2(v + \omega - 2)}{v(\omega - 2)^2(\omega - 4)}, \quad \omega > 4. \tag{24}$$

See Forbes et al. (2011), chapter 20.

To obtain the mean and variance of $\lambda$ we must first substitute $\omega = N - k - 1$ in **Error! Reference source not found.** and **Error! Reference source not found.** and then $v = k$ in **Error! Reference source not found.** in order to have the moments of the F statistic in terms of the overall sample size, $N$ and the number of covariates, $k$. To obtain the mean we then substitute again in **Error! Reference source not found.** to find

$$E[\lambda] = 1 + \frac{k}{N-k-3}, \quad k \leq N-4. \tag{25}$$

To find the variance we must multiply by the square of $k/(N-k-1)$ to obtain

$$Var[\lambda] = \frac{2k(N-3)}{(N-k-3)^2(N-k-5)}, \quad k \leq N-6. \tag{26}$$

The above formulae use an assumption of multivariate Normality of the covariates and obviously only apply approximately to other sampling schemes. We now consider the case of a single categorical covariate with $k+1$ categories. A general contingency table for treatment and covariate categories is given in Web Table 1. This contains all the information that is necessary to calculate the VIF and also to calculate the chi-square statistic for the association of covariate and treatment.

|  |  | Category |  |  |  |  | Overall |
|---|---|---|---|---|---|---|---|
|  |  | 1 | ⋯ | $j$ | ⋯ | $k+1$ |  |
| Treatment | 1 | $f_{11}$ | ⋯ | $f_{1j}$ | ⋯ | $f_{1k+1}$ | $n_1$ |
|  | 2 | $f_{21}$ | ⋯ | $f_{2j}$ | ⋯ | $f_{2k+1}$ | $n_2$ |
|  | Overall | $F_1$ | ⋯ | $F_j$ | ⋯ | $F_{k+1}$ | $N$ |

Web Table 1 Joint frequencies of values of a categorical covariate and a treatment indicator given in the form of a contingency table.

We first consider the contribution, $Chi_j^2$ of category $j, j=1\cdots k+1$ to the chi-square statistic. We have

$$Chi_j^2 = \frac{(f_{1j} - E_{1j})^2}{E_{1j}} + \frac{(f_{2j} - E_{2j})^2}{E_{2j}}, \tag{27}$$

where $E_{ij}, i=1,2$ is the expected frequency given no association. Substituting $E_{ij} = n_i F_j / (n_1 + n_2)$ we have

$$Chi_j^2 = \frac{(f_{1j} n_2 - f_{2j} n_1)^2}{n_1 n_2 F_j}. \tag{28}$$

Dividing this by the total number of observations we obtain the contribution, $R_j^2$ to the overall value of $R$ as

$$R_j^2 = \frac{Chi_j^2}{(n_1 + n_2)} = \frac{(f_{1j} n_2 - f_{2j} n_1)^2}{(n_1 + n_2) n_1 n_2 F_j}. \tag{29}$$

As regards the regression approach, we assume that we have a treatment indicator such that

$$Z = -\frac{n_1}{n_1 + n_2}, treatment = 1$$

$$Z = \frac{n_2}{n_1 + n_2}, treatment = 2 \quad . \tag{30}$$

This sums to zero and has sum of squares $\quad SS_Z = \frac{n_1 n_2}{n_1 + n_2}. \tag{31}$

Next, by dispensing with an intercept we can use $k+1$ dummy binary covariates, one for each category. Since the categories are mutually exclusive and exhaustive 1) for any given subject, exactly one of the dummy covariates is 1, all others being zero, 2) all subjects with the same category have the same dummy coding.

The predicted values for the treatment indicator can then be calculated from

$$\hat{\mathbf{Z}} = \mathbf{X}'(\mathbf{X}'\mathbf{X})^{-1} \mathbf{X}'\mathbf{Z}, \tag{32}$$

where,

$$\mathbf{X}_{N \times (k+1)}$$

is the matrix of dummy variable predictors and $\mathbf{Z}$ is the vector of treatment indicators. It then follows, that for any subject in category $j$, the prediction depends on the dummy variable associated with that category and only the treatment indicators for the subjects associated with that category are needed. Thus, sparse nature of $\mathbf{X}$ and the diagonal form of $(\mathbf{X}'\mathbf{X})^{-1}$ mean that predicted value for such a subject are given by

$$\frac{f_{2j}}{F_j} \frac{n_1}{n_1 + n_2} - \frac{f_{1j}}{F_j} \frac{n_2}{n_1 + n_2} = \frac{1}{F_j} \frac{(f_{2j} n_1 - f_{1j} n_2)}{(n_1 + n_2)}. \tag{33}$$

Thus, the total of the *squared* predicted values from all $F_j$ subjects in that category is

$$\frac{1}{F_j} \frac{(f_{2j} n_1 - f_{1j} n_2)^2}{(n_1 + n_2)^2}. \tag{34}$$

Finally, we divide **Error! Reference source not found.** by **Error! Reference source not found.** to obtain the same result as in expression **Error! Reference source not found.**, thus showing the equivalence of the two approaches.

The case of two categories is that of a binary covariate and the relationship where that is so, is noted as one of 11 different interpretations of the chi-square statistic given by (Mirkin, 2001).

The chi-square connection is a reminder that for small samples, in the case of categorical covariates the formulae for moments of the VIF may not work well. Indeed, it is not hard to see, to give a trivial example, that if we have a binary covariate and equal numbers in the two groups, so that $n_1 = n_2$, then if $F_1 = F_2$, complete confounding is possible and for small samples may not be rare. On the other hand, the chi-square connection also encourages the belief that for moderately large samples the formulae we have given may work well.